\documentstyle[aps,multicol,prl,epsf]{revtex}
\newcommand{\bleq}{\ifpreprintsty
\else
\end{multicols}\vspace*{-3.5ex}{\tiny
\noindent\begin{tabular}[t]{c|}
\parbox{0.493\hsize}{~} \\ \hline \end{tabular}}
\fi}
\newcommand{\eleq}{\ifpreprintsty
\else
{\tiny\hspace*{\fill}\begin{tabular}[t]{|c}\hline
\parbox{0.49\hsize}{~} \\
\end{tabular}}\vspace*{-2.5ex}\begin{multicols}{2}
\fi}
\newcommand{\bcols}{\ifpreprintsty\else\begin{multicols}{2}\fi}
\newcommand{\ecols}{\ifpreprintsty\else\end{multicols}\fi}

\begin{document}

\title{Theory of Transition Temperature of Magnetic Double Perovskites}
\author{A.\ Chattopadhyay$^1$ and A.\ J.\ Millis$^2$}
\address{
$^1$ Department of Physics and MRSEC, University of Maryland\\
College Park, MD 20742\\
$^2$ Center for Materials Theory\\
Department of Physics and Astronomy, Rutgers University\\
Piscataway, NJ 08854\\
}
\date{\today}
\maketitle

\widetext
\begin{abstract}
\noindent
We formulate a theory of double perovskite coumpounds such as Sr$%
_2$FeReO$_6$ and Sr$_2$FeMoO$_6$ which have attracted recent attention for
their possible uses as spin valves and sources of spin polarized electrons.
We solve the theory in the dynamical mean field approximation to
 find the
magnetic transition temperature $T_c$. We find that $T_c$ is determined by
a subtle interplay between carrier density and the Fe-Mo/Re site energy
difference, and that the non-Fe same-sublattice hopping acts to reduce
$T_c$. Our results suggest that presently existing materials do not optimize 
$T_c$.

\end{abstract}



\bcols

Identification of a ferromagnet with high spin polarization at room
temperature and stable surface properties is an important goal in the field
of magnetic materials. Such a system would allow, for example the
fabrication of 'spin-valve' devices of greatly improved efficiency for
magnetic field sensing\cite{spinvalve}, the development of new magnetic
recording media\cite{Black00}, and perhaps the construction of improved
sources of spin-polarized electrons for 'spin-tronic' applications\cite
{spintronic}. One promising family of materials are the 'double perovskites'
\cite{Sleight72,Tokura98,Prellier00}. These are compounds of chemical formula
$ABB^{\prime }O_{6}$, with $A$ an alkaline earth such as $Sr$, $Ca$ or $Ba$,
and $B,B^{\prime }$ two different transition metal ions.  Double perovskites
in which $B$ is $Fe$ and $B^{\prime }$ is $Mo$ or $Re$ are of particular
recent interest because they seem\cite{Tokura98} to be metallic ferrimagnets
with very high magnetic transition temperatures and highly spin-polarized
conduction bands. However, neither the physics nor the materials science of
these compounds is yet well understood. The magnetic transition temperature
and whether the ground state is metallic or insulating vary as A is changed
from Ba to Sr to Ca \cite{Gopal00}. Mis-site ($B-B^\prime$) disorder has a
pronounced effect\cite{Ogale99}.

In this paper we take a step towards a theoretical
understanding for these
materials. We derive a many-body Hamiltonian, using band theory
calculations\cite{Tokura98} to fix important parameters. We calculate the
magnetic transition
temperature $T_{c}$, and determine how different material
parameters affect
it. Our results should provide guidance in attempts to
design double perovskite materials with improved properties, and an
appropriate starting
point for calculations of other properties.

Double perovskites have a crystal structure which generalizes the
familiar
$ABO_{3}$ perovskite structure by having two B-site ions, which in the ideal
structure alternate in a simple
two sublattice pattern. The band theory has
been determined\cite{Tokura98}. The conduction bands are derived from
transition metal
$B$-site $t_{2g}$ d-orbitals, in agreement with quantum
chemical considerations\cite{Gopal00}. There are six conduction bands per
spin
 direction per unit cell; roughly, one triplet arises mainly from
the
 $d_{xy,xz,yz}$ orbitals on the $Fe$ and the other from the
same orbitals on Mo/Fe. The occupied bands are fully polarized
at $T=0$.

Because the near-fermi-surface bands are derived from transition
metal
d-orbitals, we argue that a simple tight-binding parametrization of the
band
theory is adequate. We therefore model the ideal compound as a cubic
lattice
of transition metal sites, divided in the usual way into two
interpenetrating
 sublattices, which we denote as A(Mo/Re) and B(Fe).  On each site 
we include 3 $t_{2g}$ orbitals,
$d_{xy},d_{xz},d_{yz}$. A good fit to the band structure requires
 both first
and second neighbor hoppings, probably because the Mo/Re conduction 
electrons come from the $4d$ shell, which is spatially extended. If 
only first and second neighbour hoppings are considered, then the $t_{2g}$ 
orbital symmetry implies that a
given orbital can mix only with
orbitals of the same symmetry and only with
orbitals in the appropriate plane
($d_{xy}$ couples only to $d_{xy}$ orbitals
in the $xy$ plane).

The band theory is thus a sum of three two
dimensional tight binding models, which we write as $H_{band}= H_{on-site} +
H_{xy}^{hop} + H_{yz}^{hop} + H_{xz}^{hop}$, with $a^{\dagger}_{im\sigma}$
creates an electron on site $i$ of orbital state $m$ and spin $\sigma$.
The on-site term $H_{on-site}$ consists of an $A-B$ site energy difference, 
which has a spin-independent term  $\Delta_m$ (independent of orbital $m$ in
cubic symmetry) and a spin dependent term $J\sigma$. 
\bleq
\begin{equation}
H_{xy}^{hop}=\sum_{p,\sigma }\left(
\begin{array}{cc}
a_{p(xy)\sigma }^{\dagger} & b_{p(xy)\sigma }^{\dagger}
\end{array}
\right)
\left(
\begin{array}{cc}
-4t_{aa}cos(p_{x})cos(p_{y}) & -t_{ab}\left[ cos(p_{x})+cos(p_{y})\right]
\\
-t_{ab}\left[ cos(p_{x})+cos(p_{y})\right] & -4t_{bb}cos(p_{x})cos(p_{y})
\end{array}
\right)
\left(
\begin{array}{c}
a_{p(xy)\sigma } \\
b_{p(xy)\sigma }
\end{array}
\right)
\label{xyhop}
\end{equation}
\eleq

We now fix parameters by computing the density of states (shown in
Fig.~\ref{dmftdos}) and comparing to band calculations. The calculated
tight-binding band structure displays features that depend sensitively on
parameters. The parameters which best fit ref.\cite{Tokura98}
are shown in Fig.~\ref{dmftdos}, with uncertainties of $10-15\%$ on all
quantities.

\begin{minipage}{3.29in}
\begin{figure}[htbp]
\epsfxsize=3.3in
\epsfysize=3.0in
\epsffile{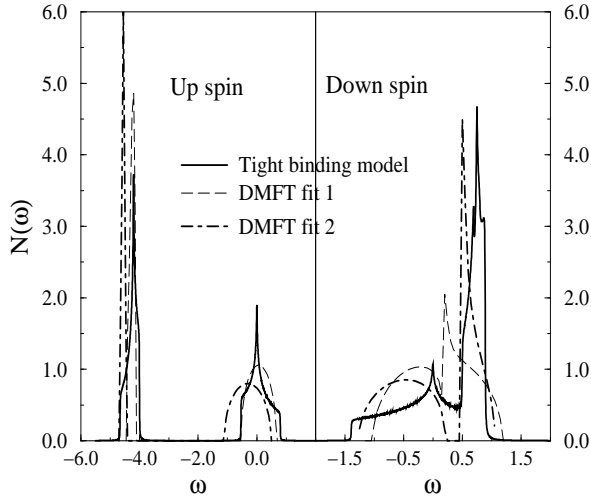}
\vspace{0.1in}
\caption{{ Density of states from tight-binding model and dynamical
 mean field theory. The solid line shows the tight-binding DOS
($t_{ab}=0.5$ eV, $t_{aa}=0.18$ eV, $t_{bb}=0.05$ eV,
$\Delta=2.35$ eV, $J=1.85$ eV); DMFT fit-1 is the local DOS
from DMFT for essentially the same parameter values as the
tight-binding model ($W_{AB}=0.5$ eV,$W_{AA}=0.22$eV, $W_{BB}=0.05$ eV,
$J=2.2$ eV and $\Delta=2.0$ eV). DMFT fit-2 is another fit to
the tight-binding model ( $W_{AB}=0.4$ eV,$W_{AA}=0.4$eV,
$W_{BB}=0.02$ eV, $J=2.5$ eV and $\Delta=2.0$ eV ). The first DMFT fit
gives a $T_c$ that is twice as much as the second one. }}
\label{dmftdos}
\end{figure}
\end{minipage}
\vspace{0.1in}

For the $|t_{aa}| < |t_{ab}|$ case of relevance here the band minima and
maxima are at $\frac{1}{2}\left[(\Delta-J) -4(t_{aa}+t_{bb})\right] \mp
\frac{1}{2}\sqrt{\left[4(t_{aa}-t_{bb})+(\Delta-J)\right]^{2}+4t_{ab}^{2}}$.
As $\omega$ is increased through $\Delta -J$, the density of states has
a step up. For $t_{bb}=0$, the hybridization vanishes along a line
in $\vec{k}$-space which intersects the van-Hove point, causing the
density of states $N(\omega)$ to diverge as
$N(\omega\approx \Delta -J) \sim (\omega-\omega^*)^{-1/2}
\ln \left[1/(\omega-\omega^*) \right]$ as $\omega\rightarrow \omega^*=
\Delta -J$ from above. A nonzero $t_{bb}$ eliminates the square-root 
divergence and moves the van-Hove singularity away from $\omega=\Delta-J$.
Thus the leading edge of the sharp peak in the density of states fixes
$\Delta-J$ and the upper and lower band edges determine $t_{ab}$ and
$t_{aa}$. Changing $t_{bb}$ from $0.05eV$ to $0$ changes the best fit
$t_{ab}$ from $0.5eV$ to $0.4eV$, and cause the density of states in the
$\omega \sim \Delta$ region to be dominated by the $(\ln
\omega)/\sqrt{\omega}$ singularity, making it asymmetric (unlike the
published band structure calculation \cite{Tokura98}). For later use we
note that the "$t_{aa}$-only" bandwidth is slightly less than the
"$t_{ab}$-only" bandwidth.

The band structure is strongly spin-polarized. In the majority spin
sector, one band (dominantly $B$ i.e. $Fe$ states) is full and the other
band is empty; In the minority spin sector one band (dominantly $A$, i.e.
$Re/Mo$) is partly filled and the other ($B$) is empty. In other words,
the B-site adopts a fully polarized filled-shell configuration, and motion
of A-carriers onto B sites depends on the spin, providing the connection
between carrier motion and spin alignment which favors ferromagnetism.

One sees from ref.~\cite{Tokura98} that in the majority-spin sector the
A-derived states form a much narrower band than do the A-derived
states in the minority sector. This shows that virtual processes which
take the B-site from $d^5$ to $d^4$ state are much less important than
those which take the B-site from $d^5$ to $d^6$. Further,
Hund's rules indicate that a $d^4$ or $d^6$ state of less than maximal 
$Fe$-spin must be even less favorable than a maximal spin $Fe$
$d^4$ configuration. Therefore, the local physics on the B($Fe$) site 
is described by
\begin{equation}
H_{loc}^B =\sum_{i\in B,a\alpha \beta }
b_{ia\alpha }^{\dagger} (-\Delta + J\hat{S}_{i}\cdot \vec{\sigma }_{\alpha
\beta })b_{ia\beta }
\end{equation}
where $\hat{S}$ is a unit vector representing the direction of the spins on
the $Fe$. Here we have chosen the zero of energy to be the A-level
energy and have restricted to cubic symmetry. Our approximation treats
the $d^5$-maximal spin $d^6$ energy splitting correctly but is only
an approximate representation of the unimportant higher energy states.
We further argue that the small filling and high degree
of spin polarization of the A( Mo/Re ) site means that we may neglect
electron-electron interactions on this sublattice. Thus we propose
the many body Hamiltonian
\begin{equation}
H=H_{xy}^{hop}+H_{xz}^{hop}+H_{yz}^{hop}+ H_{loc}^B
\end{equation}

We solve this model via the dynamical mean field
approximation
(DMFT)\cite{Georges96}. This method has been widely applied to
models, such
as the Hubbard or Kondo-lattice model, with one atom per unit cell.
In this situation, one formulates a local problem specified
by the local
action $S_{loc}=\Sigma _{\omega }g(\omega )c^{\dagger}(\omega)
c(\omega
)+H_{int}$, where $g(\omega )$ is a mean field function to be determined
by a
self-consistency condition. To generalize this
structure to the situation
of present interest we introduce two mean field
functions, ${\bf a}(\omega )$ and ${\bf b}(\omega )$ for the A and B
sublattices respectively. We use boldface to denote
tensors
depending on spin and orbital indices. Our choice of interaction
Hamiltonian
allows us to integrate over the fermion fields, yielding
\begin{equation}
S_{loc}=Tr\ln [{\bf a}]+Tr\ln [{\bf b}+ \Delta +J\vec{S}\cdot \vec{\sigma }]
\end{equation}
Here the trace is over the frequency, spin and orbital indices, and an
average over the orientation of the core spin must still be performed. We
now write the self-consistency conditions. Because an electron on site A may
hop to either another A-site ion or to another B-site ion but the spin and
orbital indices are conserved by the band part of $H$, we require three
couplings ($W_{AA}$ connecting $A$ site to $A$ bath; $W_{AB}$ connecting $A$
site to $B$ bath and conversely, and $W_{BB}$ connecting $B$ site to $B$
bath).  We define the local Green functions 
${\bf G}_{a} =\langle {\bf a}^{-1}\rangle$ and 
${\bf G}_{{\bf b}} =\left\langle ({\bf b}+\Delta + J\vec{S}\cdot 
\overrightarrow{{\bf \sigma }})^{-1}\right\rangle$ ($\langle \rangle $ 
means average over the partition function constructed from
S$_{loc}$) and obtain
\begin{eqnarray}
{\bf a} &=& \omega +\mu - W_{AA}^{2}{\bf G}_{{\bf a}}-W_{AB}^{2}
{\bf G}_{{\bf b}} \\
{\bf b} &=& \omega +\mu - W_{AB}^{2}{\bf G}_{{\bf a}}-W_{BB}^{2}{\bf G}_{
{\bf b}}
\label{dmft}
\end{eqnarray}

These equations are the generalization to the present case of the
widely-used semicircular density of states equations
\cite{Georges96,Chattopadhyay00}. They differ from the
equations successfully
 used to describe CMR manganites\cite{Chattopadhyay00,Furukawa} in that the
'core
spin' interaction occurs only on the B-site. Note that because $H^{hop}$
is
diagonal in the orbital index, these equations are diagonal as well.
However, the
partition function which determines the spin average involves all
orbitals.
We shall consider only ferrimagnetic states with magnetization
direction
specified by a unit vector $\hat{m}$, for which it is convenient to
express
the spin dependence as $a_{\alpha \beta }(\omega )=a_{0}
(\omega)+a_{1}(\omega )\hat{m}\cdot \vec{\sigma}_{\alpha \beta }$ (and
similarly
for ${\bf b}$).

We fix parameters by comparing the calculated $T=0$ DMFT density of
states to the tight-binding one. Our fits are shown in Fig.~\ref{dmftdos};
we are able to reproduce the general structure, including the upper and lower
band edges reasonably well, but there are some differences of detail. The DMFT
always exhibits a square root singularity at $\omega = J- \Delta$. Because
this is integrable, and indeed is above the chemical potential, it is not
important for our subsequent  considerations.  The DMFT is more likely to
exhibit a gap in the density of states for $\omega$ slightly smaller than
$J-\Delta$. In the band theory, the $A-B$ hybridization  vanishes  along lines
in $k$-space ( for $H_{xy}^{hop}$, this is along  $cos(p_x)+cos(p_y)=0$ ); this
feature is absent in DMFT. Finally, van Hove  singularities are present in the
band theory but not in DMFT. The "$W_{AA}$-only" $T=0$ bandwidth
is $4W_{AA}$, the "$W_{AB}$-only" bandwidth is $4W_{AB}$; comparison to the
tight-binding model suggests $W_{AB} \sim 2 W_{AA}$.

\begin{minipage}{3.29in}
\begin{figure}[htbp] 
\epsfxsize=3.3in 
\epsfysize=3.0in 
\epsffile{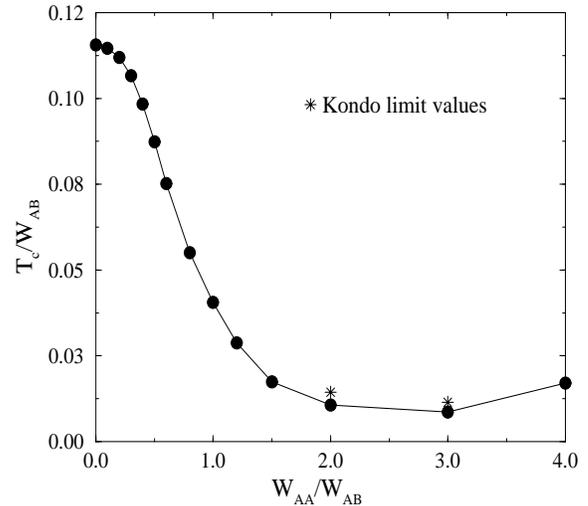} 
\vspace{0.1in} 
\caption{{ Variation of magnetic $T_c$ with the ratio of the hops 
$W_{AA}/W_{AB}$, as calculated from Eq.(~\ref{Tcgen}) using 
the "best-fit" value $(J-\Delta)/W_{AB}=0.4; n=1$. Band theory suggests 
$W_{AA}/W_{AB}\approx 0.6$. The values of $T_c$ calculated in the Kondo 
limit described in the text are shown as stars. The upturn at 
$W_{AA}/W_{AB}> 3$ arises because the A-band begins to mix with the 
majority B-states. 
}} 
\label{Tc} 
\end{figure} 
\end{minipage}
\vspace{0.1in}

We now evaluate the magnetic transition temperature, $T_c$, which we define as
the temperature at which the paramagnetic solution becomes linearly unstable
to a ferrimagnetic one. The spin structure of the mean field parameters is
expressed by the non-magnetic ($a_0$) and the magnetic ($a_1$) components,
and at $T_c$ nonzero values of $a_1,b_1$ become possible. Linearizing the
equations in $a_{1},b_{1}$ we obtain for $T_c$:
\bleq
\begin{equation}
1 = - \sum_{n}2 J^2
\frac{W_{AB}^{4}+(a_{0}^{2}-W_{AA}^{2})W_{BB}^{2}}
{(a_{0}^{2}-W_{AA}^{2})\left[\left((b_0+\Delta)^2 - J^2\right)^2 - \left(
(b_0+\Delta)^2 - \frac{1}{3} J^2\right) W_{BB}^2\right] - W_{AB}^4
\left((b_0+\Delta)^2 - \frac{1}{3}J^2\right) }
\label{Tcgen}
\end{equation}
\eleq

Our two different DMFT fits to the tight-binding model (Fig.~\ref{dmftdos}),
yield $T_c$'s of $495K$(DMFT fit 1) and $200K$(DMFT fit 2), for the
band filling appropriate for Mo.
Eq.(~\ref{Tcgen}) is a mean-field result.
Our previous experience
comparing DMFT to Monte Carlo results for
double-exchange models shows that the mean-field expression overestimates
$T_c$ by about
 25$\%$\cite{Chattopadhyay00}, implying physical $T_c$'s of
$370$ and $150K$
for these two fits. It is interesting that $T_c$'s are
comparable to or larger than $T_c$'s observed in the manganese perovskites.
We attribute this to the higher orbital degeneracy (which allows each 
and to be partially filled, favoring ferromagnetism) and very strong 
Fe-Mo/Re overlap.

We see that the calculated $T_c$'s depend sensitively on parameters. To
elucidate this, we study this dependence in more detail.
We begin with the effects of varying $W_{AA}$ shown in Fig.~\ref{Tc}, 
considering first the limit $W_{AA}\gg W_{AB}$. In this case one has a wide
band of $A$ site
carriers, weakly coupled to a spin-polarized "resonant level" on the
$B$-site, a situation familiar from heavy-fermion physics\cite{Hewson}. 
Known results for this problem imply a carrier-spin coupling 
$I \sim W_{AB}^2/(J-\Delta)$ and a $T_c \sim I^2/W_{AA}$. The DMFT 
equations can be solved in this limit, leading to precisely this 
behavior (shown as stars in Fig.~\ref{Tc}). Thus $T_c$ is maximized at
$W_{AA}=0$.

\begin{minipage}{3.29in}
\begin{figure}[htbp] 
\epsfxsize=3.3in 
\epsfysize=3.0in 
\epsffile{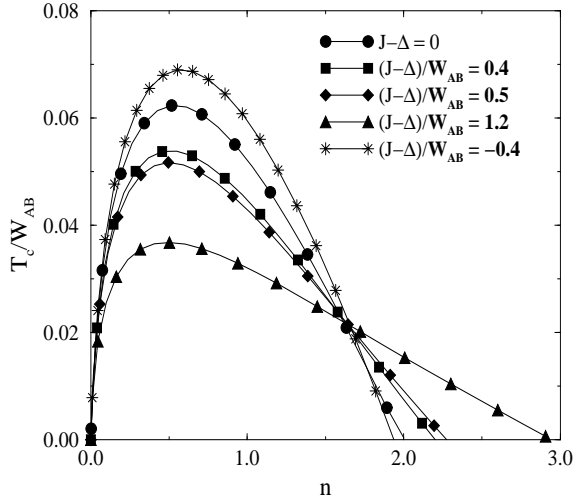} 
\vspace{0.1in} 
\caption{{ Variation of magnetic $T_c$ with band filling $n$, in the
$(\Delta+J)\rightarrow \infty$ limit, for different values of $J-\Delta$ and
$W_{AA}/W_{AB}=0.6; W_{BB}=0$.}}  
\label{Tcdel} 
\end{figure} 
\end{minipage}
\vspace{0.1in}

We now turn to the effects of varying the Fe-$d^6$/Mo energy difference
$J-\Delta$, and the carrier density, $n$. We measure the carrier density
as number added beyond the Fe $d^5$ configuration; thus Mo corresponds
to $n=1$ and Re to $n=2$. Numerical results are shown in Fig.~\ref{Tcdel};
a complicated non-monotonic behaviour is evident, which may be understood
by consideration of the simplifying limits $W_{AA}\rightarrow 0$ and
$(\Delta +J)\gg W_{AB}$. In this "double-perovskite-double-exchange" limit
carrier motion is strongly constrained by the need to have carrier and
core spins parallel. Now specialize further to the limit
$|J-\Delta|\gg W_{AB}$, so that the carriers are essentially confined to
one sublattice, with effective bandwidth $W_{AB}^2/|J-\Delta|$. The
model then maps onto the standard double-exchange one, with three orbitals
per cell and Hund's coupling $J+\Delta$. The standard arguments
\cite{Chattopadhyay00} then show that at low $n$ one has a ferromagnet
with a $T_c$ determined by bandwidth (so increasing as $|J-\Delta|$
decreases), while as $n\rightarrow n_{orb}$ (one electron per orbital) 
some other non-ferromagnetic (probably incommensurate
cf \cite{Chattopadhyay00})  arrangement of spins becomes
favorable. The density range over which a non-ferromagnetic ground
state is preferred increases as $J-\Delta$ decreases, essentially
because the greater effective bandwidth increases the energy gain from
populating 'wrong-spin' orbitals, and this accounts for the 
decrease in the ferromagnetic $T_c$ with decreasing $|J-\Delta|$ at larger n.
This trend is in qualitative agreement with the variation in $T_c$ on
changing Mo ($n=1$) for Re (n=2) in Ba and Sr-based double perovskites
\cite{Prellier00,Maignan99,Galasso66}.

To summarize, we have formulated a many-body Hamiltonian which contains
the essential physics of the Fe-based double perovskite compounds, and have
used it to determine the factors affecting the ferro(or ferri)-magnetic
transition temperature. Same-sublattice (Mo-Mo) hopping is not small in
the actual materials, and acts to lower $T_c$. The higher band fillings
of the Re compounds make it more difficult to realize high ferri-magnetic
$T_c$s, essentially because o fother competing forms of magnetic order.
We therefore suggest that a search for higher $T_c$ materials focus on
$4d^1$ systems and on ways to reduce the same sublattice hopping, as well
as more to closely match the Fe $d^6$ and B$^\prime$ site energies. On the
other hand, systems based on $4d^2$ ions are more likely to exhibit
interesting many-body physics and non-trivial ground states.

The Hamiltonian and approximation scheme we have proposed allows
a number of immediate extensions. From the tight-binding band-theory and
the DMFT self energy, the dc and optical conductivities may be calculated.
The effects of a lattice distortion which lifts the degeneracy of the three
$t_{2g}$ orbitals \cite{Gopal00} can be computed by allowing the site energy
$\Delta$ to depend on orbital index. Finally, the "cavity field"
interpretation of dynamical mean field theory allows a straightforward
investigation of the effects of mis-site ( Re-Fe interchange) disorder. Work
in all of these directions is in progress.

We thank S.\ B.\ Ogale, J.\ Gopalakrishnan and S.\ W.\ Cheong
for helpful discussions and the University of Maryland MRSEC and
NSF-DMR-9705482 for support.

\ecols

\end{document}